\documentclass[useAMS,usenatbib,letter]{mn2e}
\usepackage{epsfig} 
\usepackage{graphicx}
\usepackage{amssymb}
\usepackage{natbib}
\usepackage{times}
\def\int{{\em INTEGRAL}}
\def\XMM{{\em XMM-Newton}}

\def\chan{{\em Chandra}}
\def\swift{{\em Swift}}

\def\j16479{IGR\,J16479-4514}

\def\igrj11{IGR\,J11215-5952}

\def\2s{2S\,0114+65}
 
\begin{document}

\title[IGR\,J16479-4514 in eclipse]{IGR\,J16479-4514: the first eclipsing supergiant fast X-ray transient?}

\author[E. Bozzo, et al.]
{E. Bozzo$^{1,2}$\thanks{email: bozzo@oa-roma.inaf.it},   
L. Stella, $^{1}$ 
G. Israel, $^{1}$ 
M. Falanga,$^{3}$ 
S. Campana,$^{4}$\\
$^{1}$INAF - Osservatorio Astronomico di Roma, Via Frascati 33,
00044 Rome, Italy. \\
$^{2}$Dipartimento di Fisica - Universit\`a di Roma Tor Vergata, via
della Ricerca Scientifica 1, 00133 Rome, Italy.\\ 
$^{3}$CEA Saclay, DSM/IRFU/Service d'Astrophysique (CNRS FRE 2591), F-91191, Gif sur Yvette, France.\\
$^{4}$INAF – Osservatorio Astronomico di Brera, via Emilio Bianchi 46, I-23807 Merate (LC), Italy.\\
}

\date{Received 2008 August 8; accepted 2008 September 19.}

\maketitle

\begin{abstract}
Supergiant fast X-ray transients are a new class of 
high mass X-ray binaries recently discovered with \int.\ 
Hours long outbursts from these sources have been observed on numerous occasions at luminosities of 
$\sim$10$^{36}$-10$^{37}$~erg~s$^{-1}$, whereas  
their low level activity at $\sim$10$^{32}$-10$^{34}$~erg~s$^{-1}$ has not been deeply investigated 
yet due to the paucity of long pointed  observations with high sensitivity X-ray telescopes.
Here we report on the first long ($\sim$32~ks) pointed \XMM\ observation of 
\j16479,\ a member of this new class. This observation was carried out in March 2008, shortly after an 
outburst from this source, with the main goal of investigating 
its low level emission and physical mechanisms that drive the source activity.  
Results from the timing, spectral and spatial analysis of the EPIC-PN \XMM\ observation show 
that the X-ray source \j16479\ underwent an episode of sudden obscuration, possibly an 
X-ray eclipse by the supergiant companion.  
We also found evidence for a soft X-ray extended halo around the source that is most readily 
interpreted as due to scattering by dust along the line of sight to \j16479.\ 
We discuss this result in the context of the gated accretion scenarios   
that have been proposed to interpret the behaviour of supergiant fast X-ray transient. 
\end{abstract}

\begin{keywords}
X-rays: binaries - binaries: eclipsing - stars: individual
  (\j16479) -stars: neutron - X-rays: stars
\end{keywords}

\section{Introduction}
\label{sec:intro}

Supergiant Fast X-ray transients (SFXTs) are a new class of 
high mass X-ray binaries (HMXBs), recently discovered with \int.\  
These sources display sporadic outbursts lasting from minutes to hours  
with peak luminosities of $\sim$10$^{36}$-10$^{37}$~erg~s$^{-1}$, 
and spend long time intervals at lower X-ray luminosities, 
ranging from $\sim$10$^{33}$~erg~s$^{-1}$ to 
$\sim$10$^{34}$~erg~s$^{-1}$. 
In some SFXTs a very faint state, with a typical 
luminosity of $\sim$10$^{32}$~erg~s$^{-1}$ has also been observed.  
The outbursts of SFXTs have so far been studied in relatively fairly detail, whereas 
 the lower luminosity 
states remained poorly known due to the paucity of 
long pointings with high sensitivity X-ray telescopes.  
Data collected with both wide field instruments, such as \int,\ and imaging X-ray telescopes,  
such as those on board \chan,\ \swift\ and \XMM,\ showed that the outburst spectra are well described 
by a power law of photon index $\Gamma$$\simeq$1-2 and an absorption column density that is larger than the 
interstellar Galactic value \citep[$N_{\rm H}$$\sim$10$^{22}$-10$^{23}$~cm$^{-2}$,][]{sguera06,walter06}. 
Similar spectral properties were inferred at luminosities of $\sim$10$^{33}$-10$^{34}$~erg~s$^{-1}$ 
\citep{walter06,sidoli08}, whereas indications were found that very faint states might be characterized 
by significantly softer spectra \citep{zand05}.  
The short duration of SFXT outbursts likely indicates the accretion flow towards the NS is not mediated by a disk (viscous
timescales are of order of weeks to months). Instead, proposed models generally involve accretion onto a neutron star 
(NS) immersed in the clumpy wind of its supergiant companion \citep{zand05,leyder,walter07,bozzo08}, and suggest these 
sources have orbital periods $\gtrsim$7-10~days (or, equivalently, orbital separations 
$\gtrsim$2.5~R$_*$, with R$_*$ the radius of the supergiant star).  \\
Different models make different predictions with respect to properties and origin of states with luminosities  
below the outburst levels.   
Observations of these states, however, are still sparse and the properties of SFXTs at low luminosities 
are poorly known at present. It is generally believed that the X-ray luminosity in these states is powered by residual 
accretion onto the NS, taking place at much reduced rate than in outburst.  
In fact, large variations in the mass accretion rates in SFXTs are expected due to: i) a centrifugal and/or magnetic gating mechanism 
acting on a moderately clumpy wind \citep{bozzo08}; 
ii) clumps in the wind of the supergiant companion with extreme velocity and/or density 
contrasts \citep{walter07}; iii) the presence of two components with different densities 
in the wind of the supergiant companion \citep{sidoli07}. According to interpretation (i), the wide dynamic range 
in the X-ray luminosity (up to $\sim$5 decades) displayed by SFXTs, if accompanied by a slow spin period ($\gg$100~s),  
might indicate that these sources host ``magnetars'' \citep[i.e., neutron stars with extremely high magnetic 
fields, $\sim$10$^{15}$G,][]{duncan92}. 
During very faint states, an important contribution to the X-ray luminosity 
may also derive from shocks in the strong wind of the OB supergiant star \citep{zand05}.

Here we report on the first long high sensitivity observation ($\sim$32~ks) of the SFXT 
\j16479,\ when the source was in a low activity state. 
This source was detected in outbursts many times and appears to be characterized by a persistent 
luminosity of $\sim$10$^{34}$~erg~s$^{-1}$ \citep{sguera08,romano08,walter07}. 
The \XMM\ target of opportunity observation was carried out after \swift\ discovered   
a very bright outburst from this source on March 19 \citep{romano08b}, and was aimed at  
investigating its low level emission and gaining insight in the physical 
mechanisms that drive its activity.

\section{Data analysis and results}
\label{sec:observation}

\begin{figure}
\centering
\includegraphics[scale=0.33,angle=-90]{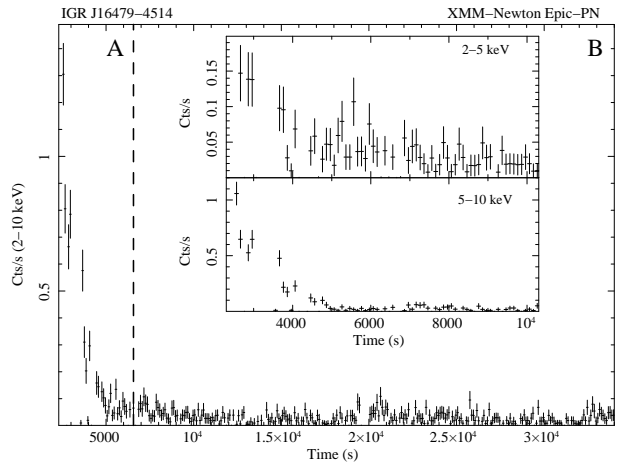}
\caption{EPIC-PN light curve of \j16479\ in  the 2-10 keV band. The bin time is 100 s 
and the start time is March 21 16:42:42UT. 
During the first 4~ks of the observation the source flux decayed from a higher  
(``A'') to a lower (``B'') level. 
The insert in the figure shows the first 8~ks of the observation in the  
2-5~keV and 5-10~keV energy bands.} 
\label{fig:lcurve} 
\end{figure}

\XMM\ \citep{j01} observed \j16479\ between March 21 14:40:00UT and March 22 01:30:00UT 
for a total time span of $\sim$37~ks. The EPIC-PN and EPIC-MOS cameras were 
operated in the small-window and full frame mode, respectively.     
We processed the observation data files (ODFs) by using the standard \XMM\ Science Analysis
System (SAS~7.1). The total effective exposure time was $\sim$32~ks; the remaining observing time 
was discarded due to high radiation from solar activity. 
The EPIC-MOS data were strongly contaminated by single reflections photons from a nearby bright source 
(likely 4U\,1642-45), and thus we report in the following on the analysis 
of the EPIC-PN data only.  
The extracted images, spectra and light curves were analyzed by using 
{\sc Heasoft} version 6.4. 
Figure~\ref{fig:lcurve} shows the 2-10~keV EPIC-PN light curve of the observation:  
in the first $\sim$4~ks \j16479\ was caught during the decay from a higher (``A'')
to a lower (``B'') flux state. The latter extended for most of the observation ($\sim$28~ks).  
We used a bin time of 1000~s in all the light curve fits; all errors are 90\% confidence level. 
The mean count rate during state B was 0.024$\pm$0.002,  
i.e., a factor $\sim$50 lower than the maximum count rate during state A.  
In order to search for spectral changes, we extracted light curves in the 
2-5~keV and 5-10~keV bands. These are shown in the insert in Fig.~\ref{fig:lcurve} 
(first 8~ks of the observation). A fit with an exponential model plus a constant revealed a much 
slower decay in the soft band light curve: the e-folding time was  
5400$\pm$1500~s and 910$\pm$100~s, in the 2-5~keV and 5-10~keV bands, respectively. 

\begin{figure}
\centering
\includegraphics[scale=0.35,angle=-90]{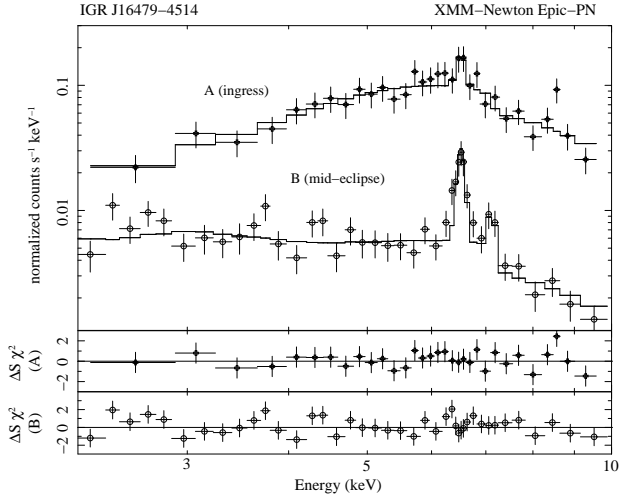}
\caption{2-10~keV spectra extracted from intervals A  
and B of Fig.~\ref{fig:lcurve}. 
The best fit models and the residuals from these fits 
are also shown (see Sect.~\ref{sec:observation}).}
\label{fig:spectra}
\end{figure}

The energy spectra of states A and B are shown in Fig.~\ref{fig:spectra}, and   
were rebinned in order to have at least 20 counts per bin.   
Fits to these spectra were tried first by assuming an absorbed power law plus a 
Gaussian line at $\sim$6.4~keV. 
This model provided a good fit to the spectrum of state A (reduced $\chi^2$=1.1)
for a flatter power law index (0.4$^{+0.5}_{-0.4}$) than previously measured for 
the source \citep[1.6$\pm$0.2,][]{sidoli08}). 
The absorption column density was marginally larger 
(1.2$^{+0.6}_{-0.3}$ $\times$10$^{23}$~cm$^{-2}$ 
as opposed to $\sim$7$\times$10$^{22}$~cm$^{-2}$).  
During state B this simple model gave a poorer fit (reduced $\chi^2$=1.7), 
a much reduced column density 
(4.0$^{0.6}_{-0.7}$ $\times$10$^{22}$~cm$^{-2}$) and a power law index compatible 
with that of state A.  The equivalent width (EW) of the Fe fluorescence line 
at $\sim 6.4$~keV showed a striking increase from $\sim260$~eV (state A) to 
$\sim$900~eV (state B). This indicated that in state B (most of) the 
X-ray flux from the source was obscured along our line of sight, but 
kept on shining on the Fe-line emitting material. 
The shape of the light curve (see Fig.~\ref{fig:lcurve}) and evolution of 
the spectrum across the state A-state B transition 
presented also remarkable similarities with the eclipse ingress 
of eclipsing X-ray sources such as
OAO\,1657-415 \citep{audley06}. Moreover the slower decay of the 
soft X-ray light curve (in turn similar to that observed in OAO\,1657-415)
suggested that \j16479\ is seen through an 
extended dust-scattering halo \citep[see e.g.,][]{day91,nagase92}. 

To  confirm this we analysed the radial distribution of the X-ray photons
detected from \j16479\ and compared it with the
point spread function (PSF) of the \XMM\ telescope/EPIC-PN camera 
(HEW of $\sim$12.5''). We did this 
separately in the 2-5~keV and 5-10~keV bands 
for states A and B. The results are shown in Fig.~\ref{fig:psf}.
The radial distribution of photons from \j16479\ was compatible
with the \XMM\ PSF for both energy ranges during state A and for
the 5-10~keV range during state B. On the contrary the 2-5~keV
source photons during state B had a considerably more extended distribution than the \XMM\ PSF. 
This behaviour is precisely the one expected from
a quickly obscured source seen through a dust scattering halo: soft X-ray 
photon emitted when the source is unobscured are scattered along 
our line of sight by interstellar dust, reaching us 
also after X-ray source obscuration, as a results of the 
longer path that they follow. The dust scattering halo is most prominent 
at low energies (due to the energy dependence of the scattering cross
section) and when the source is obscured, because contamination 
by direct photons from the source is virtually absent. 
For a halo size $\theta$ of tens of arcsec radius 
(see the top-right panel of Fig.~3) and the estimated source distance of 4.9~kpc \citep{rahoui08}, 
the longer path followed by scattered photons gives rise to a delay \citep[see e.g.,][]{thompson08}   
\begin{equation}
\delta t=5.3 d_{\rm 4.9kpc} \left(\theta/{\rm arcsec}\right)^2 x/(1-x)\sim5300~s, 
\end{equation}
where $x$ is the fractional distance of the halo from the source, $d_{\rm 4.9kpc}$ is the source 
distance in unit of 4.9~kpc, and we used $\theta$=30'' and $x$=1/2. 
The typical delay is thus of few hours, in agreement with 
the e-folding decay time of the soft X-ray light curve in state B. 

Motivated by the above findings  
we considered the spectral model that has been used in studies
of eclipsing X-ray binaries seen through a dust-scattering halo, that is 
\citep{audley06,ebisawa96} 
\begin{eqnarray}
I(E) & = & e^{\sigma(E) N_{\rm H}}[I_{\rm s} E^{-\alpha}+e^{\sigma(E) N_{\rm Hw}}I_{\rm w} E^{-\alpha} + \label{eq:xspec} \\ 
&& e^{\sigma(E) N_{\rm Hd}}I_{\rm d} E^{-\beta} + I_{\rm ln1} e^{-(E-E_{\rm ln1})^2/(2\sigma_{\rm ln1}^2)} + \nonumber \\ 
&& I_{\rm ln2} e^{-(E-E_{\rm ln2})^2/(2\sigma_{\rm ln2}^2)}] \nonumber
\end{eqnarray}
Here the continuum consists of three components seen through an interstellar 
column density of  $N_{\rm H}$: two  power laws with ''the same`` photon index 
$\alpha$ (normalizations $I_{\rm s}$ and $I_{\rm w}$), and a third power law with photon 
index $\beta$=$\alpha$+2 (normalization $I_{\rm d}$). 
The two power laws with the same photon index represent respectively, the source emission that 
is received directly at the earth (i.e., the direct component), and the source emission that is 
scattered along our line of sight by material in the immediate surroundings 
of the X-ray source, likely the wind from the massive companion star 
(we term this ``wind scattered component''). The scattered component will preserve the 
slope of the incident spectrum, while photoelectric absorption by the wind material  
can in principle reduce the number of low energy photons that are scattered, thus  
mimicking a higher column density in the scattered component. This is why the 
wind scattering term in Eq.~\ref{eq:xspec} comprises an extra absorption component 
($N_{\rm Hw}$). 
Iron features (around $\sim 6.4$~keV and $\sim 7.0$~keV see Fig.~2) are represented 
by the Gaussians in Eq.~\ref{eq:xspec} 
and arise also from reprocessing of the source radiation in the wind.  
The direct component is expected to dominate the X-ray emission when the source 
is out of eclipse. As the NS approaches the companion's limb, 
the source photons directly reaching us will propagate 
through denser and denser regions of the companion's wind, such that
increased photoelectric absorption is to be expected shortly before the eclipse 
ingress ($N_{\rm Hd}$$\gtrsim$$N_{\rm Hw}$).   
The third power law component, with photon index $\beta$=$\alpha$+2, originates from 
small angle scattering of (mainly) direct photons off interstellar dust grains 
along the line of sight \citep{day91,nagase92}; therefore, it should be characterized 
by a similar column density to that of the direct component away from eclipse. 
The dust scattering cross section decreases steeply for increasing photon energies, such that the 
spectrum of the scattered photons is softened by a factor of $E^{-2}$. 
Both the wind and dust scattered components are expected to be most prominent during the eclipse,  
when the direct component is occulted by the companion star.   
\begin{figure}
\centering
\includegraphics[scale=0.35,angle=-90]{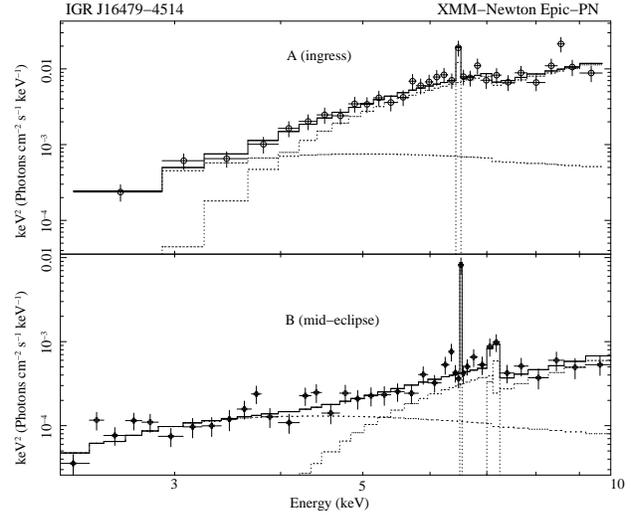}
\caption{2-10~keV unfolded spectra of Fig.~\ref{fig:spectra}. 
The best fit model components are displayed with dotted lines.} 
\label{fig:unfolded}
\end{figure} 

Owing to limited statistics and degeneracy in some the of spectral fit 
parameters, it proved impossible to disentangle the contribution from the two power laws of index
$\alpha$ in the \XMM\ spectra of \j16479.\ 
In practice we fitted a model with only two continuum components, a single power law 
of slope $\alpha$ (with normalisation $I_{\alpha}$ and column density $N_{H,\alpha}$) 
and a power law of slope $\beta$ (with normalisation $I_{\beta}$ and column density $N_{H,\beta}$). 
The former component was taken to approximate the sum of the two $\alpha$-slope power laws 
in Eq.~\ref{eq:xspec}, with the direct component 
dominating before the eclipse ingress and the wind scattered component dominating 
during the eclipse. The relevant model was thus 
\begin{eqnarray}
I(E) & = & e^{\sigma(E) N_{\rm H,\beta}}I_{\beta} E^{-\beta} + 
e^{\sigma(E) N_{H,\alpha}} [ I_{\alpha} E^{-\alpha} + \nonumber \\ 
&& I_{\rm ln1}e^{-(E-E_{\rm ln1})^2/(2\sigma_{\rm ln1}^2)} + \nonumber \\
&& I_{\rm ln2} e^{-(E-E_{\rm ln2})^2/(2\sigma_{\rm ln2}^2)} ]
\label{eq:ccc}. 
\end{eqnarray} 
We kept $\alpha$ fixed  at 0.98, the best fit value found by \citet{romano08} 
when the source was in outburst 
(and thus away from an X-ray eclipse); consequently we fixed also $\beta$=2.98.  
The best fit models are shown in Fig.~\ref{fig:spectra}, and the corresponding parameters 
are given in Table~\ref{tab:fit}. In Fig.~\ref{fig:unfolded} we plot also the unfolded spectra. 
While in state A, a power law component with slope $\alpha$ plus a Gaussian line at 
$\sim 6.5$~keV (EW of $\sim 190$~eV) provided a reasonably good fit also for a 
fixed value of $\alpha\sim 0.98$ ($\chi^2$/d.o.f.=33/28). However, adding 
the dust scattered component significantly improved the fit ($\chi^2$/d.o.f.=20.4/26). 
In state A the relatively high source flux and small EW of the iron fluorescence line 
testifies that (most of) the emission is likely due to the direct component. 
In order to investigate this, we also accumulated spectra during the first 1.5~ks and last 2.5~ks of state A 
(not shown in Fig.~\ref{fig:spectra}). 
These two spectra were reasonably fit ($\chi^2$/d.o.f.=18.6/16, 13.3/9) by assuming an absorbed 
powerlaw with fixed photon index 0.98 and an absorption column density of  
30$^{8}_{-7}$$\times$10$^{22}$~cm$^{-2}$ and 5.6$^{3.8}_{-3.0}$$\times$10$^{22}$~cm$^{-2}$, respectively  
(the addition of a Gaussian line at $\sim$6.5~keV did not significantly improved any of these fits). 
We interpreted the decreasing absorption column density revealed by these fits as the effect 
of a change in the dominating spectral component during the early ingress to the eclipse. 
Specifically, the direct component dominates the first part of the state A and is absorbed by 
the very dense companion's limb; instead, most of the X-ray emission during the last 2.5~ks of this state 
is likely due to the wind scattered component, which is in turn absorbed by a lower dense material, 
in agreement with the expectation $N_{\rm Hd}$$\gtrsim$$N_{\rm Hw}$ discussed above. 

In state B, the ratio $I_{\alpha}$/$I_{\beta}$ is larger than the corresponding value  
obtained during state A, indicating a dust scattering halo whose intensity decreases
slowly during the eclipse. The intensity of the power law of slope $\alpha$ 
decreased substantially, while the EW of the Fe-line at $\sim 6.5$~keV increased to $\sim$770~eV. 
We also found evidence ($\sim$2$\sigma$) for an additional Fe-line at 
$\sim 7.1$~keV with a $\sim$300~eV EW, consistent with being the $K_{\beta}$. 
Owing to the poor statistics of the present observation, 
we were unable to investigate further the nature of this line; however, we note that the  
ratio between the normalizations of the $K_{\alpha}$ and $K_{\beta}$ lines, as inferred from fits 
in Table~\ref{tab:fit}, is roughly in agreement with the expected value $\sim$0.15-0.16         
\citep[see e.g.,][]{molendi03,ibarra07}. 
The most natural interpretation of this is that in state B  
the direct emission component is occulted along our line of sight, while the spectrum 
we observe is the sum of a dust scattered component, dominating at lower energies, and a 
wind scattered component characterised by a high absorption. 
The marked increase in the EW of the Fe-line at $\sim$6.5 across the state A-state B 
transition testifies that the region where the line is emitted is larger 
than the occulting body (the supergiant companion, if we are dealing with an eclipse); this provides strong evidence that 
the uneclipsed emission in state B at hard X-ray energies (the $\alpha$-slope power law) arises mostly 
from photons scattered by the wind in the immediate surrounding of the source.  
In both states A and B the measured $N_{H,\beta}$ is consistent with the absorption column density reported  
by \citet{romano08}; this provides further support in favour of the $\beta$-slope power law representing 
the outburst emission seen through a dust scattering halo with a delay of $\gtrsim$1~hour. 
These findings and interpretation are in line with those that emerged from detailed 
studies of some persistent eclipsing HMXBs. 

We obtained also spectral fit in which $\alpha$ 
was allowed to vary as free parameters, while $\beta$=$\alpha$+2.  
However, this did not lead to a significant improvement of any of the fits, while the best 
fit parameters were consistent with the ones found above. 
\begin{table}
\centering
\caption{Best fit parameters of the spectra during states A and B. $F_{\rm 2-10~keV}$ is the absorbed 
flux in the 2-10~keV band, and the X-ray luminosity $L_{\rm X}$ is derived from this flux assuming a 
source distance of 4.9~kpc \citep{rahoui08}. Errors are at 90\% confidence level.} 
\begin{tabular}{lll}
\hline
\hline
\noalign{\smallskip} 
FIT PARAMETERS & STATE ``A'' & STATE ``B''\\
\hline
\noalign{\smallskip} 
$N_{H,\alpha}$ (10$^{22}$~cm$^{-2}$) & 35$^{18}_{-13}$ & 54$^{26}_{-25}$\\
\noalign{\smallskip}
$N_{H,\beta}$ (10$^{22}$~cm$^{-2}$) & 9$^{-9}_{6}$ & 6$^{1}_{-2}$ \\
\noalign{\smallskip}
$I_{\alpha}$ & 1.8$^{0.4}_{-0.3}$$\times$10$^{-3}$ & 1.1$^{0.5}_{-0.4}$$\times$10$^{-4}$\\
\noalign{\smallskip}
$\alpha$ & 0.98 (frozen) & 0.98 (frozen) \\
$I_{\beta}$ & 5$^{8}_{-4}$$\times$10$^{-3}$ & 8$^{2}_{-3}$$\times$10$^{-4}$\\
\noalign{\smallskip}
$\beta$ & 2.98 (frozen)& 2.98 (frozen)\\
\noalign{\smallskip}
$E_{\rm ln1}$ (keV) & 6.53$^{0.06}_{-0.07}$& 6.51$^{0.03}_{-0.02}$\\
\noalign{\smallskip}
$I_{\rm ln1}$ & 5$^{3}_{-3}$$\times$10$^{-5}$ & 1.8$^{1.0}_{-0.4}$$\times$10$^{-5}$\\
\noalign{\smallskip}
$EW_{\rm ln1}$ (keV) & 0.15 & 0.77 \\
\noalign{\smallskip}
$E_{\rm ln2}$ & - & 7.11$^{0.06}_{-0.09}$\\
\noalign{\smallskip}
$I_{\rm ln2}$ & - & 8$^{7}_{-4}$$\times$10$^{-6}$\\
\noalign{\smallskip}
$EW_{\rm ln2}$ (keV) & - & 0.28 \\
\noalign{\smallskip}
$\chi^2$/(d.o.f.) & 20.4/26 & 36.5/33 \\
\noalign{\smallskip}
$F_{\rm 2-10~keV}$ & 1.0$\times$10$^2$ & 7.5 \\
(10$^{-13}$~erg~cm$^{-2}$~s$^{-1}$) & & \\
\noalign{\smallskip}
$L_{\rm X}$ (10$^{33}$~erg~s$^{-1}$) & 29.0 & 2.2 \\
\noalign{\smallskip}
\hline
\hline
\noalign{\smallskip}
\end{tabular}
\label{tab:fit}
\end{table} 

No evidence for a coherent modulation was found in the present data up to periods 
of $\sim$1000~s (3$\sigma$ upper limit in the 80\%-90\% pulse fraction\footnote{Here for pulse  
fraction we mean the ratio between the pulsed emission and the total emission from the NS.} range).

\section{Discussion and Conclusion}
\label{sec:discussions} 

We reported on the first long ($\sim$32~ks) pointed observation of \j16479\ during 
a period of low X-ray emission. 
The decay of the light curve and the increase of the iron line EW between states A and B, 
likely indicated that the X-ray radiation from this source was being obscured by an occulting body.  
Based on the present knowledge of \j16479,\ we suggested that \XMM\ might have caught the source entering  
an X-ray eclipse. The possibility that the obscuration event was caused by a wind clump cannot be excluded 
at present. However, we regard this interpretation as unlikely, because photoelectric absorption would be 
expected to cause a more pronounced flux decrease at soft X-ray energies than at hard X-ray energies 
(at least in the early stages of the obscuration event) contrary to the results presented here. 
On the other hand, an energy-independent obscuration could be caused by a fully ionized clump 
that is optically thick to electron scattering, a very unlikely possibility in consideration 
of the low X-ray luminosity of the source.   
Obscuration by a tilted accretion disk, which is known to cause similar 
effects in other HMXBs \citep[see e.g,][]{zane04}, also does not appear as a very promising possibility here.     
This is because the outbursts of \j16479,\ and SFXT in general, have durations that are incompatible with 
viscous timescales in a typical accretion disc (see Sect.~\ref{sec:intro}).  

In order to interpret the source flux and spectral variations revealed by the \XMM\  
observation, we considered the same scenario that has been extensively discussed for 
eclipsing HMXBs with a dust scattering halo \citep[see e.g.,][]{audley06,ebisawa96}:
during the eclipse ingress source photons 
reaching us directly dominated the X-ray flux we observed; while in eclipse the residual 
X-ray flux is due to the sum of a wind-scattered component and a dust scattering halo.
Strong evidence for the wind scattered component in the case of \j16479\ comes 
from the marked increase of the EW of the iron fluorescence line at $\sim 6.5$~keV. 
A tens of arcsec wide dust-scattering halo around \j16479\ could be directly seen in the soft X-ray 
\XMM\ images accumulated during the eclipse. Moreover the residual soft X-ray 
decreased on a timescale of $\sim 1$~hr after the eclipse onset, as expected
for a dust scattering halo of the size seen by \XMM\ at the source distance of 4.9~kpc. 
\begin{figure}
\centering
\includegraphics[scale=0.48]{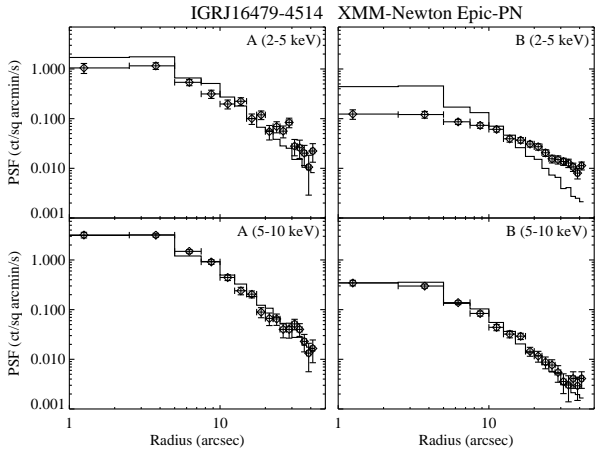}
\caption{\XMM\ EPIC-PN PSF as a function of the distance from the source. The upper panels are for 
images extracted in states A and B in the energy range 2-5~keV. 
Instead, the PSF in the lower panels are extracted from the same images but in the 5-10~keV  
energy range. For this analysis we used the {\sc ximage} tool (version 4.4).} 
\label{fig:psf}
\end{figure}   

\begin{figure}
\centering
\includegraphics[scale=0.5]{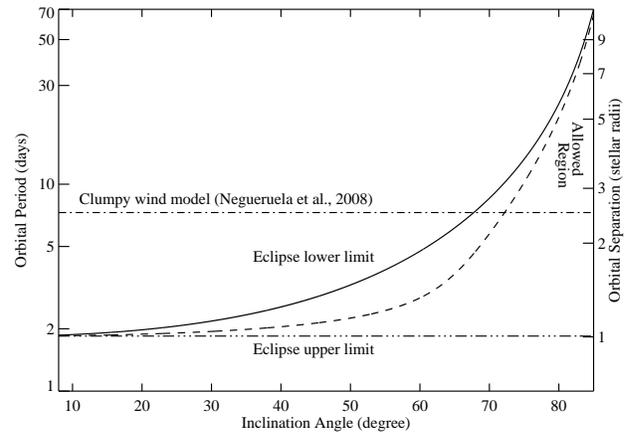}
\caption{Orbital period constraints on \j16479,\ as a function of the binary inclination angle. 
The solid line is given by Eq.~\ref{eq:relation} assuming an eclipse duration of 
28~ks, the dashed line is for an eclipse duration of 55~ks, and the triple-dot-dashed 
line is for an eclipse lasting half of the binary orbital period. 
The dot-dashed line is the lower limit on the orbital period obtained by applying the clumpy 
wind model to \j16479.\ According to this model, the region of allowed parameters in this plot 
indicates \j16479\ is a high-inclination SFXT.}
\label{fig:eclipse}
\end{figure}

We conclude that \j16479\ is the first SFXT 
that displayed evidence for an X-ray eclipse. 
In this case, some constraints on the orbital period of this system can be derived by using 
\citep{rappaport83} 
\begin{equation}
R_*=a\left[\cos{i}^2+\sin{i}^2\sin{\Theta_{\rm e}}\right]^{1/2}.  
\label{eq:relation}
\end{equation}
Here $i$ is the inclination of the orbit to the plane of the sky, $\Theta_{\rm e}$ is the 
eclipse half-angle, $a$ is the binary separation, and we assume in the following 
a supergiant star with a mass of $\sim$30~M$_{\odot}$ and a radius of R$_*$$\sim$20~R$_{\odot}$. 
We consider circular orbits, as it is generally expected for an HMXB with 
a supergiant companion \citep{rappaport83}. 
By using Kepler's law, Eq.~\ref{eq:relation} can be solved for the orbital 
period, as a function of the inclination angle, for any fixed value of the eclipse duration. 
This is shown in Fig.~\ref{fig:eclipse}. The solid line in this figure represents 
the solution obtained by assuming an eclipse duration of $\sim$28~ks. This value 
should be regarded as a lower limit on $\Theta_{\rm e}$, since in the \XMM\ observation 
discussed here there is no evidence for the eclipse egress. 
The dashed line is for an eclipse that lasts for 55~ks (chosen for comparison with the previous case), 
whereas the triple-dot-dashed line is for an eclipse that extends for half of the binary orbital period.   
The latter provides an upper limit  on $\Theta_{\rm e}$. 
We also include in the figure, the lower limit on the orbital period that was obtained by 
\citet{negueruela08}, by assuming the sporadic outbursting activity of SFXTs is related 
to accretion of clumps from the supergiant stellar wind (see Sect.~\ref{sec:intro}). 
If we consider that the clumpy wind model applies to \j16479,\ then the region of allowed parameter 
space in Fig.~\ref{fig:eclipse} suggest this source is a high inclination SFXT. 
From this figure we also note that, for this source, an eclipse with a duration much 
longer than the $\sim$28~ks observed here is very unlikely: in fact the longer is the duration of the eclipse, 
the smaller is the allowed parameter region in Fig.~\ref{fig:eclipse}. In particular, 
a much longer eclipse would require a very 
high inclination angle and short orbital period that can be hardly reconciled with the requirements of the clumpy 
wind model (the NS orbit should be very close to the supergiant surface, and a persistent luminosity 
of $\gtrsim$10$^{36}$~erg~s$^{-1}$ would be expected). 

\citet{walter07} derived the average quiescent flux of \j16479\ as observed by \int\ over 
$\sim$67~d, which converts to an X-ray luminosity of few 10$^{34}$~erg~s$^{-1}$. 
This relatively high luminosity is comparable with that of state A, and thus provides  
support in favor of state B being an unfrequent state, as expected in the eclipse interpretation 
(eclipses cannot last more than half of the orbital period).   
 
At present, it is not clear whether eclipses should be expected also in other SFXTs. 
However, we note that, if other SFXTs display X-ray eclipses, their maximum luminosity swing  
might not be due to genuine changes in the mass accretion rate,  
impacting on the application of the gating accretion model for SFXT \citep{bozzo08}.   
Further observations of \j16479,\ as well as other SFXTs in quiescence, will improve our 
knowledge of the low level emission of these sources, 
and will help clarifying this issue.

\section*{Acknowledgments}
We thank Norbert Schartel and the XMM-Newton staff, 
for carrying out this ToO observation, and the referee, Roland Walter,  
for useful comments and suggestions. 
EB thanks L. Sidoli and P. Romano for their collaboration 
during the early stages of this work.  
EB also thanks E. Piconcelli for useful 
discussions. This work was partially supported 
through ASI and MIUR grants.

\end{document}